\documentclass[twocolumn]{revtex4-1}

\usepackage{amsmath,amssymb,amsfonts,epsfig,graphicx,float,dcolumn,bm}

\begin{document}


\title{A Self-Assembled Metamaterial for Lamb Waves}

\author{A. Khanolkar$^{1}$, S. Wallen$^{1}$, M. Abi Ghanem$^{1}$, J. Jenks$^{1}$, N. Vogel$^{2}$, and  N. Boechler}
\affiliation{
$^1$ Department of Mechanical Engineering, University of Washington, Seattle, WA, 98195 USA \\
$^2$ Institute of Particle Technology and Cluster of Excellence Engineering of Advanced Materials, Friedrich Alexander-University, Erlangen-Nuremberg, 91058 Germany \\
}

\begin{abstract}
We report the design and characterization of a self-assembled, locally resonant acoustic metamaterial for Lamb waves, composed of a monolayer of $1.02$ $\mu$m polystyrene microspheres adhered to a $1.3$ $\mu$m thick free-standing silicon membrane. A laser-induced transient grating technique is used to generate Lamb waves in the metamaterial and measure its acoustic response. The measurements reveal a microsphere contact resonance and the lowest frequency spheroidal microsphere resonance. The measured dispersion curves show hybridization of flexural Lamb waves with the microsphere contact resonance. We compare the measured dispersion with an analytical model using the contact resonance frequency as a single fitting parameter, and find that it well describes the observed hybridization. Results from this study can lead to an improved understanding of microscale contact mechanics and to the design of new types of acoustic metamaterials.\end{abstract}

\maketitle

Locally resonant acoustic metamaterials are a type of composite material that have been the subject of intense study over the past fifteen years \cite{LeamyRev} due to their ability to exhibit extreme \cite{Lemoult2012}, anisotropic \cite{Sheng2011}, negative \cite{Liu2000}, strongly absorbing \cite {Sheng2012}, and locally-tailorable \cite{Fang2011} effective properties. These unique properties stem from the interaction of propagating acoustic waves with subwavelength resonant elements forming the composite \cite{NoteonSuprawavelength}. This can present fabrication challenges as wavelengths are reduced, including challenges involved with fabricating the resonators themselves, and in fabricating large areas of composite containing such complex microstructure. In this respect, colloidal self-assembly is a promising solution, as it has been shown to enable the simple, inexpensive, and fast fabrication of complex, ordered structures composed of nano- or microscale elements in one to three dimensions \cite{SelfAssembly1}. These advantages have driven the use of self-assembly strategies in multiple areas, particularly in the design of phononic crystals \cite{ThomasReview,Fytas2006}, photonic crystals \cite{SelfAssemblyPhotonic,ThomasReview}, plasmonic sensors and nanostructures \cite{SelfAssemblyPlasmonic}, and surfaces with tailored wettability \cite{SelfAssemblyWetting}. Despite this wide use, there remain few examples of locally resonant acoustic metamaterials fabricated using self-assembly techniques \cite{Fytas2008, Boechler_PRL}. 

Lamb waves are a type of acoustic waveguide mode that occur in thin elastic plates and membranes \cite{Lamb1917} that play an important role in nanomechanical resonator \cite{Wu2011} and sensing applications \cite{Yantchev2013}. They have also been utilized in studies of sub-THz phonon transport with implications for the understanding of nanoscale thermal phenomena \cite{Cuffe2012}. Recently, several examples of locally resonant metamaterials for Lamb waves have been explored in both theoretical \cite{Pennec2008} and experimental \cite{Wu2009} settings. However, in each of the experimentally realized cases, the resonant elements forming the metamaterial had frequencies ranging from the audible regime to a few megahertz and were fabricated using conventional machining or microfabrication techniques.  
\begin{figure}[t]
\begin{center}
\includegraphics[width=8cm]{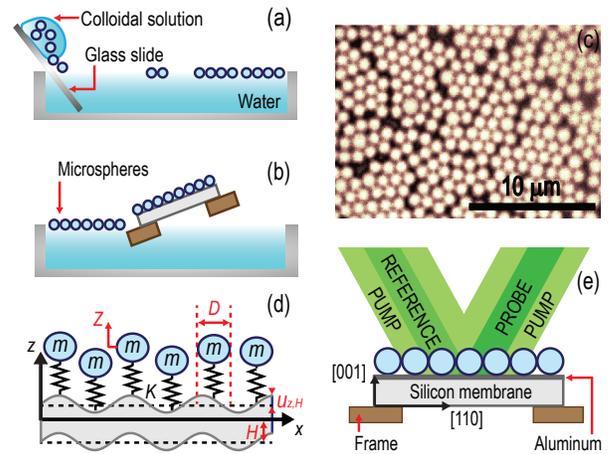}
\end{center}
\caption{\label{Figure1} Illustration of the metamaterial self-assembly fabrication procedure, including: (a) monolayer formation at the air/water interface; and (b) transfer of the monolayer to the membrane. (c) Representative image of the polystyrene microsphere monolayer. Schematic illustrations of the (d) theoretical model and (e) experimental setup. }
\end{figure} 

In this work, we present the first realization of a self-assembled, locally resonant metamaterial for Lamb waves. The metamaterial is composed of a self-assembled monolayer of microspheres adhered to the surface of a thin silicon membrane. As a result of the adhesion, the microspheres have a contact resonance where the microsphere oscillates like a rigid body, with a localized region of elastic deformation around the point of contact with the substrate. In our system, the contact resonance plays the role of the metamaterial locally resonant element and has a frequency of $200$ MHz. This type of metamaterial may have potential future advantages for acoustic wave tailoring applications, as arrays of macroscale spherical particles have been shown to support unique nonlinear dynamical phenomena due to the Hertzian relationship \cite{Hertz} between spherical elastic particles in contact \cite{NesterenkoBook,GranularCrystalReviewChapter}. To characterize the metamaterial, we use a laser-induced transient grating technique \cite{RogersReview2000,NelsonTG2012} to excite long-wavelength (relative to the particle size) Lamb waves in the metamaterial and measure its acoustic response. This approach was recently used to study the interaction of microsphere contact resonances with Rayleigh surface acoustic waves (SAWs) \cite{Boechler_PRL}. In the dispersion curves derived from our measurements, we identify both the presence of the lowest frequency microsphere spheroidal resonance and hybridization phenomena typical of locally resonant acoustic metamaterials, in the form of a previously unobserved phenomenon: the hybridization of fundamental flexural ($A_0$) Lamb wave modes with a contact resonance of the microspheres. We develop a model that assumes an axial contact resonance and yields an analytical expression for the dispersion relation. In the calculated dispersion curves, we observe that the microsphere contact resonance not only couples with $A_0$ modes, but also with the fundamental dilatational ($S_0$) Lamb wave modes, which results in coupling between flexural and dilatational modes. We find good agreement between our measurements and the model using the microsphere contact resonance as a single fitting parameter, and compare the fitted contact resonance with predictions based on microscale contact models. 

The metamaterial is composed of a monolayer of $D=1.02$ $\mu$m diameter polystyrene microspheres deposited on, and adhered to, the aluminum side of an  aluminum-coated ($100$) silicon membrane of thickness $2H=1.3~\mu$m, as shown in Fig.~\ref{Figure1}. The thickness of the membrane was measured with ellipsometry. The aluminum film is $50$~nm thick and serves as a medium to absorb the optical pump light. To fabricate the metamaterial, we utilize a self-assembly procedure, in which polystyrene microspheres are assembled at the air/water interface and then transferred to the membrane, as shown in Fig.~\ref{Figure1}(a,b) \cite{Vogel2011}. The resulting monolayer covers nearly the entire area of the membrane, which has dimensions of $4.8$~mm $\times$ $4.8$~mm. A representative microscope image of the monolayer packing is shown in Fig.~\ref{Figure1}(c). 

The laser-induced transient grating technique~\cite{RogersReview2000,NelsonTG2012} used in this study is summarized as follows. The configuration of the beams and the sample is illustrated in Fig.~\ref{Figure1}(e). Two optical pump beams derived from a pulsed laser source (532 nm wavelength, 430 ps pulse duration, and 1 kHz repetition rate) are overlapped at the aluminum layer in the metamaterial, and form a periodic interference pattern. The pump spot has a $500$~$\mu$m diameter at $1/e^2$ intensity level. Absorption of the pump pulse light by the metamaterial induces a rapid thermoelastic expansion, which induces counter-propagating acoustic waves with a wavelength equal to the optical interference pattern that is defined by the crossing angle of the beams. The acoustic wavelength is controlled by changing a phase mask pattern used to split the pump beam into $+/-1$ diffraction orders. The detection of the metamaterial acoustic response is accomplished with a quasi-cw probe beam (wavelength $514$~nm and average power $10.7$~mW at the sample) chopped to $50$~$\mu$s pulses with an electro-optic modulator. The probe beam passes through the same set of optics as the pump beam and is focused at the center of the interference pattern to a spot of diameter $300$~$\mu$m. Pump-induced surface ripples and refractive index variations caused by the propagating acoustic waves (including contributions from the silicon membrane, aluminum film, and microspheres) lead to a time-dependent diffraction of the probe beam. The diffracted probe light is superimposed with an attenuated reference beam and is directed to a photodiode where it is recorded with an oscilloscope. 

Using this technique, we measured the acoustic response of the aluminum-coated membrane before the microspheres were deposited (the ``without-spheres'' case). In this case, the pump pulses have an energy of 19 $\mu$J and both pump and probe beams  are incident on the aluminum-silicon interface in the metamaterial, such that they pass through the silicon membrane. As a result, the measured signal for the without-spheres case includes contributions from surface ripples from both the aluminum film and the silicon membrane, and refractive index variations in the silicon. After the microsphere deposition, we measured the acoustic response of the metamaterial (the ``with-spheres'' case). In this case, both the pump and probe beams enter the metamaterial from the opposite side, such that they are incident on the aluminum-microsphere interface, and the pump pulse energy is reduced to 5.6 $\mu$J. Measuring the side with the microspheres provides enhanced sensitivity to microsphere motion \cite{Boechler_PRL}, in addition to signal contributions from surface ripples of the aluminum film. 
%
\begin{figure}[t]
\begin{center}
\includegraphics[width=8cm]{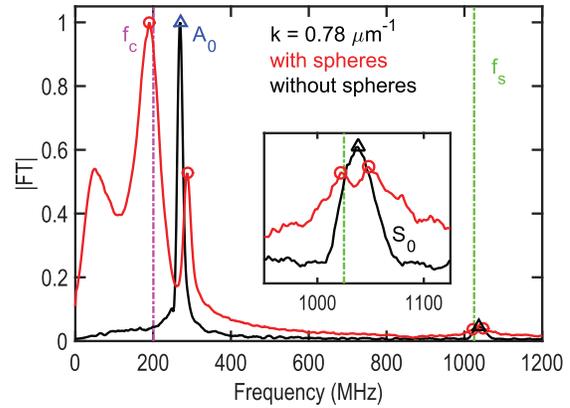}
\end{center}
\caption{\label{Figure2} Spectra of Fourier transform magnitudes where the black curve corresponds to the without-spheres case, the red curve to the with-spheres case, and each spectrum is normalized to its maximum amplitude. The markers denote the identified peaks, which are plotted also in Fig.~\ref{Figure3} using the same markers and colors. The vertical lines denote the frequencies of the fitted microsphere contact resonance and spheroidal resonance.} 
\end{figure} 

Figure~\ref{Figure2} shows the normalized Fourier spectra of the acoustic oscillations corresponding to the without- and with-spheres cases at $k=0.78$~$\mu$$\text{m}^{-1}$. In the without-spheres case, two peaks are clearly observed. We identify the low frequency peak at $270$ MHz as the $A_0$ mode of the membrane, and the second peak at $1040$ MHz, as the $S_0$ mode. Because of the presence of the aluminum film, we predominantly excite antisymmetric modes, as can be seen by the relative amplitude of the $A_0$ and $S_0$ modes \cite{Rogers1994}. We measure the dispersion of these two modes by varying the acoustic wavelength, and plotting the peaks identified in the Fourier spectra, as shown in Fig.~\ref{Figure3}. The measured spectra and their identified peaks for all wavelengths can be found in the Supplementary Information \cite{Supplementary}. Using a silicon density of \(\rho_{m} = 2.33\) g/cm\textsuperscript{3} and typical wave speeds in silicon \cite{SiliconSpeeds} of $c_L=9133$~m/s (longitudinal) and $c_T=5844$ m/s (transverse), corresponding to a propagation in the $[110]$ direction with a transverse displacement along $[001]$, we calculate the theoretical dispersion curves for the $A_0$ and $S_0$ modes in the membrane using an isotropic model \cite{Ewing}. We find good agreement between our measurements and the calculated dispersion curves as illustrated in Fig.~\ref{Figure3}. We note that the  measured frequencies are slightly lower than predicted, which we attribute to our use of an isotropic model instead of the fully anisotropic model \cite{Auld1973}. Calculations accounting for the presence of the aluminum layer showed that the variation in the $A_0$ and $S_0$ dispersion curves due to the presence of the aluminum does not exceed 1\% \cite{JonesTwoLayer1964}. A clear difference between the with- and without-spheres cases can be seen by comparing the two spectra in Fig.~\ref{Figure2}. In the with-spheres case, each of the $A_0$ and $S_0$ modes identified in the without-spheres case are surrounded by two peaks. In this case, we also see an additional peak near $70$~MHz. We observe this peak to be present in both the without- and with spheres cases for multiple wavelengths \cite{Supplementary}, and we do not consider it further for this study. 

The measured dispersion curves for the with-spheres case are plotted in Fig.~\ref{Figure3}. The with-spheres dispersion curves reveal \lq\lq{}avoided crossing\rq\rq{} \cite{Wigner1929} behavior between the $A_0$ branch and the microsphere contact resonance. The lowest frequency branch of the with-spheres dispersion curve follows the without-spheres $A_0$ branch from the origin and then diverges to approach a horizontal asymptote near the contact resonance frequency. Above the resonance, a second branch follows the $A_0$ branch at high wave vector magnitudes, but again diverges in the avoided crossing region to approach a horizontal asymptote just above the contact resonance. Measured peaks corresponding to dilatational motion closely match each other for the with- and without-spheres cases. The dispersion curves also show a flat branch at $f_s$ = 1030 MHz that intersects with the $S_0$ branch, which we attribute to microsphere spheroidal resonance. The spheroidal resonance frequency is obtained by averaging frequencies of the peaks identified along this branch. We note the presence of two peaks at $k=0.78$~$\mu$$\text{m}^{-1}$. This suggests either a hybridization between the between the $S_0$ branch and the microsphere spheroidal resonance, or simultaneous detection of the $S_0$ peak and the spheroidal resonance. 
%
\begin{figure}[h]
\begin{center}
\includegraphics[width=8cm]{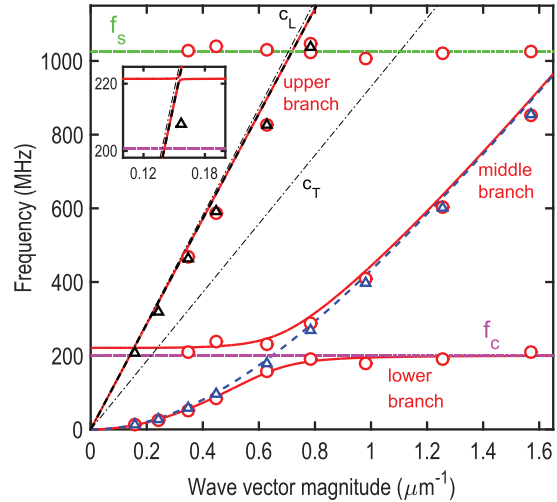}
\end{center}
\caption{\label{Figure3} Dispersion relations. The red and black markers are the measured frequency peaks for the with- and without- spheres cases, respectively. The solid red line is the dispersion calculated using our model. The dashed black line corresponds to the calculated $S_0$ modes and the dashed blue line to $A_0$ modes in the without-spheres case. The black dash-dot lines correspond to bulk waves in silicon. The horizontal lines denote the frequencies of the fitted microsphere contact resonance and identified spheroidal resonance.}
\end{figure} 

We compare the measured spheroidal resonance frequency with the frequency of the lowest frequency spheroidal mode $f(1,2)$ of a free homogeneous isotropic polystyrene sphere, where the first index denotes the mode number and the second the harmonic number. Applying the elastic equation for spheroidal modes \cite{Sato}, we calculate a frequency of $f(1,2)$ = 1030 MHz using standard longitudinal and shear velocities for polystyrene of $c_{L,p}=2350$~m/s and $c_{T,p}=1200$ m/s \cite{Fytas2006}, and a polystyrene density of $\rho_s=1.06$~g/$\text{cm}^3$ as provided by the microsphere manufacturer (Corpuscular, Inc.). As the calculated frequency is in close agreement with the measured frequency, we utilize velocities $c_{L,p}$ and $c_{T,p}$ and density $\rho_s$ to solve for the polystyrene microspheres' elastic modulus $E_s$ = $4.04$ GPa and Poisson's ratio $\nu_s$ = $0.32$. The simultaneous identification of both the spheroidal resonance and the contact resonance offers an opportunity to study the variation of each within the same microsphere array, as has previously been demonstrated for isolated nanospheres \cite{Ravaine2012}. 

To study Lamb wave propagation in our metamaterial, we adopt a similar approach as the one used previously to describe the interaction of a contact resonance of microspheres with Rayleigh SAWs \cite{Boechler_PRL}. We model the microspheres as an array of linear surface oscillators that move in the $Z$ direction, with mass $m$ and linearized normal contact stiffness $K$ attached to the top surface of the thin silicon membrane, as illustrated in Fig.~\ref{Figure1}(d). The equation of motion for the surface oscillator can be written as $m\ddot{Z}+K(Z-u_{z,H})=0$, where $u_{z,H}$ is the displacement of the substrate surface, $Z$ is the displacement of the oscillator, and the microsphere mass $m$ is calculated from the density $\rho_s$. Because the frequency of the lowest frequency spheroidal mode is much higher than the contact resonance frequency and the elastic deformation highly localized, we describe the entire mass of the microsphere moving in a rigid-body-like motion at frequencies near the contact resonance. Since the acoustic wavelength is much larger than the sphere size, we use an effective medium approach and approximate the average normal stress at the surface as the force exerted by the microsphere divided by the area of a unit cell \cite{Boechler_PRL}. By applying boundary conditions corresponding to the average normal stress at one surface of the membrane instead of stress-free boundary conditions, we obtain the following dispersion relation for $+x$-propagating flexural and dilatational Lamb waves in the metamaterial \cite{Supplementary}:
\begin{equation} \label{DR}
	\begin{aligned} 
		( \frac{\omega^2}{\omega_c^2}- 1) D_1 D_2  &= \frac{-m \omega^4q_L}{2\rho_m A k^3 c_T^4}[ \sinh{(q_L k H)}\sinh{(q_T k H)}D_1 \\
		\qquad + & \cosh{(q_L k H)}\cosh{(q_T k H)} D_2], \\
		\end{aligned}
\end{equation}
\newline	
\noindent
where $q_L = \sqrt{1 - \omega^2/(k^2 c_L^2)}$, $q_T = \sqrt{1 - \omega^2/(k^2 c_T^2)}$, \(\omega_c = 2 \pi f_c = \sqrt{K/m}\) is the angular frequency of the contact resonance, $\omega$ is the angular frequency of Lamb waves propagating through the metamaterial, $A=\sqrt{3}D^2/2$ is the area of a primitive unit cell in the hexagonally packed monolayer, and \(D_{1,2}\) are determinants whose zeros comprise the dispersion relations of uncoupled, purely flexural and dilatational modes (respectively) in a membrane with stress-free boundary conditions \cite{Ewing,Supplementary}. The term in parentheses on the left-hand side of Eq. \ref{DR} describes the resonance of the oscillators. The right-hand side is responsible for the coupling between the oscillators, and flexural and dilatational Lamb waves.

By taking the frequency of the contact resonance as the only fitting parameter, and using least squares minimization between the calculated and measured curves, we find the frequency of the contact resonance to be $f_c=200$ MHz. We plot the theoretical dispersion curves calculated using the fitted contact resonance frequency in Fig.~\ref{Figure3}. The theoretical dispersion curves show coupling between the flexural and dilatational modes due to the presence of the spheres adhered to one side of the membrane. In contrast to the coupling between Rayleigh SAWs and the microsphere contact resonance \cite{Boechler_PRL}, the middle branch (as denoted in Fig.~\ref{Figure3}) does not stop at the line corresponding to transverse polarized bulk waves, and remains non-leaky due to the confinement of the membrane. This results in a second hybridization between the middle and the $S_0$ branch. The inset in Fig.~\ref{Figure3} shows a closer view of this intersection, which is smaller than the avoided crossing near the $A_0$ branch. This demonstrates stronger coupling between flexural motion and the contact resonance than with dilatational motion. This is a result of flexural modes having large out-of-plane displacement that couples with the vertical motion of the surface oscillators, in contrast to dilatational modes with predominantly in-plane displacements. 

In Fig.~\ref{Figure2}, we observe that the largest signal contribution comes from the peak nearest the contact resonance frequency. This is the case for each of the peaks along the flat part of the lower branch, which is consistent with previous observations \cite{Boechler_PRL} and is due to large signal contributions from the microsphere motion, as the oscillator equation of motion predicts large sphere oscillations and small surface displacements near the contact resonance. We observe a different behavior along the flat part of the middle branch \cite{Supplementary}, where the peaks closest to the microsphere contact resonance are of small relative amplitude. This difference is also consistent with predictions from the oscillator equation of motion, as the flat part of the middle branch is further from the microsphere contact resonance as compared to the flat part of the lower branch. We also observe wider peak widths for peaks near the microsphere resonances despite low group velocities. Possible causes include scattering due to disorder in the monolayer and inhomogeneous broadening due to variation in the contact stiffnesses. The latter includes both particle-substrate and particle-particle stiffnesses, although at the long wavelengths measured in this study, we do not expect significant interparticle effects.

We estimate the frequency of the microsphere contact resonance using the Derjaguin-Muller-Toporov (DMT) adhesive contact model \cite{DMT1983,Boechler_PRL} and compare this estimate with the fitted contact resonance frequency. We obtain a contact resonance $f_{c,DMT} = (1/2\pi)(K_{DMT}/m)^{(1/2)}$ = 108 MHz, where $K_{DMT}$ is the stiffness of the contact. To obtain the stiffness, we linearize the DMT contact model around its equilibrium position  \cite{Supplementary}, as no anharmonic behaviors are observed for these pump powers, and we estimate the displacement of the surface to be much smaller than the equilibrium overlap of the microspheres \cite{Supplementary}. As for previous measurements of $1.08$ $\mu$m diameter silica microspheres on a thick aluminum coated fused silica substrate \cite{Boechler_PRL}, we find that the measured contact frequency is much larger than the frequency predicted utilizing the DMT contact model. The use of alternative contact models results in little variation in the predicted contact resonance frequency \cite{Supplementary}. Instead, we suggest that other uncertainties in the modeling of the contact may be the cause for the discrepancy between the estimated and the measured values of the contact resonance frequency.  

We have presented the realization of a self-assembled, locally resonant metamaterial for Lamb waves composed of a monolayer of microspheres adhered to a thin elastic membrane. With their high characteristic frequencies and small length scales, these metamaterials hold promise for the development of new types of Lamb wave-based devices. Since the response is sensitive to the state of the contact, these metamaterials may have potential applications as sensors for humidity, temperature, and micro/nanoscale material properties, and serve as a platform for the exploration of microscale contact mechanics. Because of the scalability enabled by self-assembly, this type of metamaterial may enable future studies that explore the interaction of local mechanical resonances with even higher-frequency phonon transport. Finally, the use of self-assembly approaches may also offer significant potential advantages for producing acoustic metamaterials inexpensively and on a large scale.

The authors thank J. Eliason, A. Maznev, and K. Nelson for guidance with the optical setup and for useful discussions. This work was supported by NSF grant no. CMMI-1333858 and the University of Washington (UW) Royalty Research Foundation. Part of this work was conducted at the UW NanoTech User Facility and the Washington Nanofabrication Facility, members of the NSF National Nanotechnology Infrastructure Network.

\end{document}